\documentclass[aoas,MSNbibl,nameyear,seceqn,dvips]{arximspdf}
\usepackage{dcolumn}

\doi{10.1214/11-AOAS465}
\volume{5}
\issue{3}
\pubyear{2011}
\firstpage{2109}
\lastpage{2130}

\makeatletter

\let\sv@tabnotetext\tabnotetext
  \let\sv@tabnotemark@fmt\tabnotemark@fmt
   \long\def\legend#1{{\let\tabnote@indent\leavevmode\sv@tabnotetext[]{}{#1}}}

   \def\@bmisc[#1]{%
  \get@battribute{unstr}%
  \common@pub@types%
  \let\bauthor\bbl@bauthor%
  \let\bhowpublished\@firstofone%
  \def\borganization##1{{\bauthor@style ##1}}%
}

\newcolumntype{d}[1]{D{.}{.}{#1}}
\newcommand{\p}{\mathrm{P}}
\newcommand{\E}{\mathrm{E}}
\newcommand{\var}{\operatorname{var}}
\newcommand{\Sig}{{\bolds\Sigma}}
\newcommand{\m}{{\bolds\mu}}
\newcommand{\bt}{{\bolds\beta}}
\newcommand{\1}{\mathbf{1}}
\newcommand{\Y}{{\mathbf{Y}}}
\newcommand{\X}{{\mathbf{X}}}
\newcommand{\F}{\mathbf{F}}
\newcommand{\I}{\mathbf{I}}
\newcommand{\R}{{\bolds\rho}}
\makeatother

\begin{document}
\begin{frontmatter}

\title{Generalized genetic association study with samples of related
individuals}
\runtitle{Generalized association study}

\begin{aug}
\author[A]{\fnms{Zeny} \snm{Feng}\corref{}\ead
[label=e1]{zfeng@uoguelph.ca}\thanksref{tt1}},
\author[B]{\fnms{William W. L.} \snm{Wong}\ead[label=e2]{wwl.wong@utoronto.ca}},
\author[C]{\fnms{Xin} \snm{Gao}\thanksref{tt1}\ead
[label=e3]{xingao@mathstat.yorku.ca}}
\and\\
\author[D]{\fnms{Flavio} \snm{Schenkel}\thanksref{tt1}\ead
[label=e4]{schenkel@uguelph.ca}}
\runauthor{Feng, Wong, Gao and Schenkel}
\affiliation{University of Guelph, University of Toronto, York
University and~University~of~Guelph}
\address[A]{Z. Feng\\Department of Mathematics\\ \quad and
Statistics\\
University of Guelph\\ Guelph, Ontario N1G2W1\\ Canada\\
\printead{e1}} %adresu isvedimo komanda gale!
\address[B]{W. Wong\\
Toronto Health Economics and \\ \quad Technology Assessment
Collaborative\\
University of Toronto \\
Toronto, Ontario M5S3M2\\ Canada\\
\printead{e2}}
\address[C]{X. Gao\\
Department of Matematics\\ \quad and Statistics\\
York University\\North York, Ontario M3J1P3\\ Canada\\
\printead{e3}}
\address[D]{F. Schenkel\\
Department of Animal and Poultry\hspace*{26pt}\\ \quad  Science\\
University of Guelph\\Guelph, Ontario N1G2W1\\
Canada\\
\printead{e4}}
\end{aug}
\thankstext{tt1}{Supported in part by the Natural Sciences and
Engineering Research Council of Canada (NSERC) individual discovery grant.}

% HISTORY:
\received{\smonth{3} \syear{2010}}
\revised{\smonth{2} \syear{2011}}

% ABSTRACT
%
\begin{abstract}
Genetic association study is an essential step to discover genetic
factors that are
associated with a complex trait of interest. In this paper we present a
novel generalized quasi-likelihood
score (GQLS) test that is suitable for a study with either a
quantitative trait or a binary trait. We use a logistic regression
model to link the phenotypic value
of the trait to the distribution of allelic frequencies. In our model,
the allele frequencies are treated as a response and the trait is
treated as a covariate that allows us to leave the distribution of the
trait values unspecified. Simulation studies indicate that our method
is generally more powerful in comparison with the family-based
association test (FBAT) and controls the type I error at the
desired levels. We apply our method to analyze data on Holstein cattle
for an estimated breeding value phenotype, and to analyze data from the
Collaborative Study of the Genetics of Alcoholism for alcohol
dependence. The results show a good portion of significant SNPs
and regions consistent with previous reports in the literature, and
also reveal new significant SNPs and regions
that are associated with the complex trait of interest.
\end{abstract}

% KEYWORDS
%
\begin{keyword}
\kwd{Genetic association test}
\kwd{kinship-inbreeding coefficient}
\kwd{logistic regression}
\kwd{quasi-likelihood}.
\end{keyword}

\end{frontmatter}

%s1 ###
\section{Introduction}\label{sec1}
\label{sintro}

Recent biological technology allows researchers to perform genome-wide
association studies using a dense panel of
SNPs at an affordable cost. Association studies have been widely used
to identify genome regions that are
associated with a complex trait of interest. Current methods in genetic
association
studies can be roughly categorized into two approaches: (1) studies on samples
of unrelated subjects; (2) studies on samples of related subjects, from
nuclear families, extended families,\vadjust{\eject} or from isolated/founder
populations which often include inbred individuals that are related
through multiple lines of descent.

The classical population-based association test in a case--control study
design is the simplest approach where
unrelated affected (cases) and unaffected (controls) individuals are typed.
However, for a rare disease, it is difficult to recruit independent cases
in the general population, and, more importantly, the naive analysis of
data from a general population
recruitment design may lead to false positive signals due to
confounding effects caused by the population
structure. Many researchers [\citet{Ewan}; \citet{Khoury98}; \citet{Lander94}] have
reported and discussed aspects of this
problem. For example, the confounding effect of ethnicity
is well known as the population stratification effect in the genetics
literature. For an association test with a
quantitative trait, a simple linear regression model is often used. As
noted, the association tests of
quantitative traits via population-based approaches are also subject to
the same problem of confounding by the
population stratifications.

The family-based association study design using the family based
association test
(FBAT) analysis method has become popular, as this strategy is robust
to the population heterogeneity
[\citet{Horvath}; \citet{Laird}]. In
FBAT analysis, a statistic $U$ is computed on the basis of the linear
combinations
of offsprings' genotype and phenotype expression functions. The mean
and the variance of $U$ under the null
hypothesis of no association is calculated conditional on the parental
genotype. Thus, FBAT methods typically
require the typing of family members, such as parents or siblings (for
inferring a missing parental genotype) of
each affected subject to make use of such a subject in the test. This
becomes a limitation of the method. For example,
for a late onset disease, it is difficult and sometimes impossible to
collect the information of the
family members of an affected subject. On the other hand, FBAT
typically requires heterozygous parents to compute the null
distribution of the test statistic.
Moreover, when dealing with a large pedigree, FBAT breaks down the
pedigree to small nuclear families, such that the relationship among
remotely related individuals are ignored. Similarly, FBAT does not take
into account for the relationship across related families in the analysis.
For these reasons, a family-based approach is
generally less powerful in comparison with population-based approaches
[\citet{Risch}; \citet{Bourgain}; \citet{Thornton}].

\citet{Slager} have proposed a method that was based on the Armitage
trend test with the inclusion of a variance that accounts for the
relationships among individuals from an
outbred population. However, this method cannot handle large, complex,
inbred pedigrees.
A different approach, a pedigree disequilibrium test, proposed by
\citet{Martin} can be employed to handle large
pedigree association analysis. A~founder/isolated
population-based study design has been suggested [\citet{Lander94};
\citet{Wright99}] for association mapping.
This study
design efficiently controls the confounding effect due to population
structure and has been useful
for complex trait mapping. Recently, \citet{Bourgain} proposed a
case--control association test where subjects
are sampled from a founder population with known genealogy. They
adapted the idea of a population-based
association test to test whether the allele frequencies of a specified
allele are equal between the case group
and control group, taking into account the correlations among subjects
and the inbreeding configuration within
subjects. This method can be used to analyze data from a large inbred
pedigree and is also suitable for data
from multiple pedigrees with careful control of ethnic homogeneity
[\citet{Thornton}]. The test is based on a
quasi-likelihood scoring (QLS) approach and has been shown to be more
powerful than the traditional
transmission/disequilibrium test (TDT) when samples are from
homogeneous populations.
However, these approaches are limited to binary traits.\looseness=-1

Following the line of quasi-likelihood approach proposed by \citet{Bourgain} and
\citet{Thornton} to handle the correlation structure among related subjects,
we propose a generalized linear model framework to accommodate other
types of traits. We use a logistic
regression model to link the trait to the distribution of allelic
frequencies. In our model, the observed trait of each individual is
treated as a covariate. The proportion of a
specified allele in the genotype is the response. In conventional
models, the phenotypic trait is treated as the response and the
distribution of the trait values needed to be specified. For example,
the normality assumption is often required for a quantitative trait. In
our method, the trait is treated as an explanatory variable, which
allows us to leave the distribution unspecified. On the other hand,
treating the allele frequencies of the marker as the response, we have
the exact covariance structure for the responses with the provision of
the pedigree structure or the documented genealogy. Under this
innovative modeling, we
derive the test statistic ($W_G$) and show that $W_G$ asymptotically
follows a
$\chi^2_{k-1}$ distribution, where $k$ is the number of alleles of the
marker. Our proposed GQLS test generalizes the existing approaches in
three aspects: (1) the GQLS method can establish associations between
marker's allele frequencies and all types of traits; (2) it uses a
general link function to connect the mean value of the allele frequency
with the traits; (3) our GQLS method can be extended to solve the
problem when a sample is collected from multiple subpopulations. In
this article we focus on the logistic link, but the extension
of our test to other link functions, for example, the probit function,
would be straightforward.

This paper is motivated by the challenges of analyzing data on Holstein
cattle in North America. The
aim of this study is to identify SNPs or genome regions that are
associated with
the estimated breeding values (EBVs) of a~proven bull. The EBV of a
bull predicts its genetic merit. For example, the milk yield EBV of a
bull predicts the milk yield of its female descendants. Conducting an
association study in this data set is challenging. First, dams are not
typed, and sires are typed only if they appear as proven bulls in the
data set. Thus, FBAT is not applicable to
analyze this data set. Second, most of the bulls, sires and dams, are inbred.
They are descendants from a~single complex pedigree and the
relationships among them are known but complicated.
The conventional population-based association test does not account for
this complex
relationship among subjects. Ignoring the correlation structure among
subjects would lead to an inflated positive
result. This will be shown by simulation studies in the paper.
Third, the case--control founder-population-based approach proposed by
\citet{Bourgain} is limited to binary
traits where most of the EBVs are quantitative.
Thus, the challenge of analyzing this data set becomes a
motivation for the development of our method.

We perform simulation studies on collections of pedigrees of various
sizes and on single complex pedigrees with different sizes to validate
our method. We compare the empirical performance of our method with
others. In application, we also apply our method to the Collaborative
Study of
the Genetics of Alcoholism (COGA) data provided by the Genetic Analysis
Workshop (GAW) 14 [\citet{Edenberg}; \citet{Bailey}] to demonstrate the
application in the binary trait and multiple small families study design.

The paper is organized as follows. Section~\ref{sec2} presents the proposed
generalized quasi-likelihood association test. Section~\ref{sec3} presents
the details of simulation studies to assess the validity and the power
of the proposed test compared with other methods. In Section~\ref{sec4}
applications to real data are provided to
illustrate the practical application of the proposed method.
Discussions are
provided in Section~\ref{sec5}.

%-----------------------------------------------------------------------
%Section 1
%%--------------------------------------------------------------------------
%s2 ###
\section{Methods}\label{sec2}
%s2.1 ###
\subsection{Association test with a biallelic marker}\label{sec2.1}
\label{ss1}
Suppose that in a genetic study we have a sample of
$n$ subjects that is from a single isolated/founder population or a
single pedigree. Subjects may be arbitrarily
related with a~known relationship. It is assumed that the inbreeding
configuration for each subject is also
known. Let $\X=(X_1,\ldots,X_n)'$ with $X_i$ being the phenotypic
observation of the $i$th subject. The
$X_i$ can be binary with $X_i=1$ or 0 coding for ``affected'' or
``unaffected,'' respectively, or can be continuous for a quantitative
trait. Given a biallelic
marker of interest, alleles are labeled by ``0'' and ``1.'' Let $\Y
=(Y_1,\ldots, Y_n)'$ with
$Y_i=\frac{1}{2}{}\times{}$(the number of allele 1 in subject $i$)
being the proportion of the allele 1 in the observed
genotype of subject $i$, and $Y_i=0, \frac{1}{2}$, or 1. Let $
{\bolds\mu}=(\mu_1,\ldots,\mu_n)'=\E(\Y|\X)$ that $0<\mu
_i<1$. We propose a
logistic regression model to link the expected allele frequency~$\m$
of the marker with the trait $\X$.
We let
%
%e2.1 ###
\begin{equation}
\mu_i=E(Y_i|X_i)=\frac{e^{\beta_0+\beta_1X_i}}{1+e^{\beta_0+\beta_1X_i}}.
\label{logistic}
\end{equation}
To test the association between the marker and the trait, we test
\[
H_0\dvtx  \beta_1=0       \quad \mbox{against} \quad      H_a\dvtx  \beta_1\neq0.
\]
Our model provides a natural constraint that $0<\mu_i<1$ for all $i=1,\ldots, n$.\vspace*{-1pt} Under the null hypothesis, we
have $\mu_i=\mu=\frac{e^{\beta_0}}{1+e^{\beta_0}}$ for all $i=1,\ldots, n$. The\vspace*{-1pt} mean vector of $\Y$ no longer depends on $X_i$ and becomes
${\bolds\mu}=\E(\Y)=\mu\mathbf{1}$, where $\mathbf
{1}$ is an $n$-vector of $1$'s. It can be shown that,
under $H_0$, the covariance matrix of $\Y$ is given by
${\bolds\Sigma}_0=\frac{1}{2}\mu(1-\mu)\R,$ and
%
%e2.2 ###
\begin{equation}
\R= \pmatrix{\displaystyle
1+\phi_1 & 2\phi_{12} &\cdots&2\phi_{1n}\cr\displaystyle
2\phi_{12}&1+\phi_2& \cdots&2\phi_{2n}\cr\displaystyle
\vdots&\cdots& \ddots&\vdots\cr\displaystyle
2\phi_{1n}&2\phi_{2n}& \cdots&1+\phi_n
},
\label{covmatrix}
\end{equation}
where $\phi_i$ is the inbreeding coefficient of individual $i$
and $\phi_{ij}$ is the kinship coefficient between individual $i$ and
individual $j$. See Appendix~A in the supplementary material for the
justification [\citet{Feng}]. The covariance matrix ${\bolds\Sigma}_0$ will be invertible if $\mu\neq1$ or 0, and
$\R$ is invertible provided that the monozygous twins (twins that are
genetically identical, as they
originate from a single fertilized egg) are merged and represented by
one single individual. This can be done using the multiple outputation
procedure [\citet{Follmann}].
The quasi-likelihood score function is in the form of
%
%e2.3 ###
\begin{equation}
S(\bt)=(S_{\beta_0}(\bt),S_{\beta_1}(\bt))'=D'{\bolds\Sigma}^{-1}(\Y-{\bolds\mu}),
\end{equation}
where $D$ is a $n\times2$ derivative matrix in the form of
%
%e2.4 ###
\begin{equation}
D=\frac{\partial\m}{\partial\bt}= \biggl(\frac{\partial\m
}{\partial\beta_0},\frac{\partial\m}{\partial
\beta_1} \biggr),
\end{equation}
and $\bolds\Sigma$ is the covariance matrix of $\Y$. Under the null
hypothesis, we have
${\bolds\mu}=\mu\mathbf{1}$ and the covariance matrix
$\Sig=\Sig_0$. The solution to the equation of the
quasi-likelihood score function $S_{\beta_0}(\beta_0,0)=0$ gives an
estimate of $\mu$ as
%
%e2.5 ###
\begin{equation}
\hat{\mu}=(\1'\R^{-1}\1)^{-1}\1'\R^{-1}\Y,
\label{muhat}
\end{equation}
and therefore gives the estimate of $\beta_0$ as $\hat{\beta}_0=\log
\frac{\hat{\mu}}{1-\hat{\mu}}$
under the null hypothesis. See Appendix~B in the supplementary material
for the derivation [\citet{Feng}].

When $\beta_1\neq0$, the marker is associated with the trait and
the expected value of $Y_i$ given the $X_i$ is given by equation (\ref{logistic}). For a binary trait, the two-sample model of \citet{Bourgain} in the form of
\[
\mu_i=
\cases{\displaystyle
p+r  ,&\quad if  $i$  is affected, with   $0<p+r<1 $,\cr\displaystyle
p  ,&\quad if  $i$  is unaffected, with   $0<p<1$
}
\]
becomes a special case of our model that $p=\frac{e^{\beta
_0}}{1+e^{\beta_0}}$ and $r=\frac{e^{\beta_0+\beta_1}}{1+e^{\beta
_0+\beta_1}}-\frac{e^{\beta_0}}{1+e^{\beta_0}}$. We propose a
generalized quasi-likelihood scoring statistic to test the association
between the marker and the trait. Under
the null hypothesis that $\beta_1=0$,
\[
\E[S_{\beta_1}(\beta_0,\beta_1=0)]=\E\biggl [\frac{\partial\m
}{\partial
\beta_1} \Sig ^{-1}(\Y-{\bolds\mu
}) \biggr]=0.
\]
As described by \citet{Cox}, the quasi-score statistic is given~by
%
%e2.6 ###
\begin{equation}
W=S_{\beta_1}(\hat{\beta}_0,0)'\var^{-1}_0(S_{\beta_1}(\hat{\beta
}_0,0))S_{\beta_1}(\hat{\beta}_0,0),
\label{qss}
\end{equation}
where $\hat{\beta}_0$ is the quasi-likelihood estimate of $\beta_0$
and $\var^{-1}_0(S_{\beta_1}(\hat{\beta}_0,0))$ is the $(2,2)$th
entry of the
inverse of the information matrix $\I(\bt)$ that is computed under
the null hypothesis that $\beta_1=0$.
As demonstrated by \citet{Heyde}, under the null hypothesis, $W$
follows a $\chi^2$ distribution with 1 degree of freedom
asymptotically. In our case, we obtain an explicit
expression for our generalized quasi-likelihood scoring statistic in
the form of
%
%e2.7 ###
\begin{eqnarray}
\label{wgs}
W_{\mathrm{G}}&=&\frac{2}{\hat{\mu}(1-\hat{\mu})}[\X'\R^{-1}(\Y
-\hat{\mu}\1)]'\nonumber\\
&&{}\times [\X'\R^{-1}\X-(\X'\R^{-1}\1)(\1'\R^{-1}\1)^{-1}(\1'\R^{-1}\X
)]^{-1}\\
&&{}\times[\X'\R^{-1}(\Y-\hat{\mu}\1)],\nonumber
\end{eqnarray}
where $\hat{\mu}$ is given by equation
(\ref{muhat}). See Appendix~B in the supplementary material for the
derivation [\citet{Feng}]. Note that, in equation (\ref{wgs}), we do
not need $\hat{\beta}_0$ to compute the $W_G$ statistic. $W_G$ is
expressed in a general form for both the quantitative and binary traits.
When the trait is binary, the
quasi-likelihood scoring statistic proposed by \citet{Bourgain}
becomes a special case of our $W_G$ that they are the same. Under the
null hypothesis, $W_G$ follows a $\chi^2_1$ distribution
asymptotically.

Following the same line as in \citet{Bourgain}, we generalize the
$W_G$ statistic to accommodate
$F$ independent families in an outbred
population. Among $n$ subjects, let $n_f$ be the number of subjects
that are from family~$f$ and let
$\Y_f=(Y_{1f},\ldots, Y_{n_ff})'$ be the vector of $Y$'s for subjects
that are from family $f$, $f=1,\ldots, F$.
Then, we have $n=n_1+\cdots +n_F$. Let $\Sig_f$ and~$\R_f$ be the
covariance and correlation matrix of $Y$'s for
those subjects that are from the $f$th family. If all the individuals
in the sample are outbred, the diagonal
entries of matrix $\R_f$ are equal to 1 for all $f=1,\ldots, F$. The
overall covariance matrix under the null
hypothesis is a block diagonal matrix that consists of $\Sig_1,\ldots,
\Sig_F$.
We derive that explicit form for the quasi-likelihood estimate of $\mu
$ under the null hypothesis as
%
%e2.8 ###
\begin{equation}
\hat{\mu}=\Biggl(\sum_{f=1}^F\1_f'\R^{-1}_f\1_f\Biggr)^{-1}\Biggl(\sum_{f=1}^F\1
_f'\R^{-1}_f\Y_f\Biggr),
\label{gmuhat}
\end{equation}
where $\1_f$ is the $n_f$-vector of 1's. We derive an explicit
form that
%
%e2.9 ###
\begin{equation}
W_G=\frac{2}{\hat{\mu}(1-\hat{\mu})}A'B^{-1}A,
\end{equation}
where
\begin{eqnarray*}
A&=&\sum_{f=1}^F[\X_f'\R_f^{-1}(\Y_f-\hat{\mu}\1_f)],
\\
B&=&\sum_{f=1}^F
\X_f'\R_f^{-1}\X_f-\Biggl(\sum_{f=1}^F \X_f'\R_f^{-1}\1_f\Biggr)^2\Biggl(\sum
_{f=1}^F\1_f'\R_f^{-1}\1_f\Biggr)^{-1},
\end{eqnarray*}
and $\X_f$ is the $n_f$-vector of the traits of the individuals from
the $f$th family.

%s2.2 ###
\subsection{Association test with a multiallelic marker}\label{sec2.2}
\label{ss2}
Now, suppose the marker under investigation has $k$ different alleles
and there are $n$ individuals
being sampled from a single pedigree. Let $\Y=(\Y_1',\ldots, \Y
_{k-1}')'$ be an $n(k-1)$-vector with
$\Y_j=(Y_{j1},\ldots, Y_{jn})'$ being an $n$-vector that $Y_{ji}=\frac
{1}{2}{}\times{}$(the number of allele  $j$
  in individual  $i$). Similarly to the biallelic case, we let $\m
=\E(\Y|\X)=(\m_1',\ldots,\m_{k-1}')'$ with
$\m_j=(\mu_{j1},\ldots, \mu_{jn})'$ and
\[
\mu_{ji}=\frac{e^{\beta_{0j}+\beta_{1j}X_i}}{1+\sum
_{j=1}^{k-1}e^{\beta_{0j}+\beta_{1j}X_i}}.
\]
Each random vector $2\times(Y_{1i},\ldots, Y_{k-1,i})' $ follows a
multinomial $(2, (\mu_{1i},\ldots,\break  \mu_{k-1,i})')$
distribution with $0<\mu_{ji}<1$ and $\sum_{j=1}^{k}\mu_{ji}=1$ for
all $i=1,\ldots, n$. Under the null hypothesis that the marker is not
associated with the trait, all $\beta_{1j}$'s are 0. Thus, we perform
a simultaneous hypothesis test that
\[
H_0\dvtx  \beta_{11}=\cdots =\beta_{1,k-1}=0       \quad \mbox{vs} \quad
H_a\dvtx \mbox{ at least one } \beta_{1j}\neq0, \quad  j=1,\ldots,
k-1.
\]
Here, we generalize the notation of vector $\bt$ as in the biallelic
case that $\bt=(\bt_0', \bt_1')'$ with
$\bt_0=(\beta_{01},\ldots,\beta_{0,k-1})'$ and $\bt_1=(\beta
_{11},\ldots,\beta_{1,k-1})'$. Under the null hypothesis
that $\bt_1=\mathbf{0}$, we have $\mu_{ji}=\mu_j$ for all $i$ and
rewrite the mean vector $\m=(\mu_1\1',\ldots,
\mu_{k-1}\1')'$ where $\1$ is an $n$-vector of 1's. Under the null
hypothesis, the covariance matrix of $\Y$ is
given by $\Sig=\F\otimes\R$ (the Kronecker product of matrices $\F
$ and $\R$) where $\F$ is a $(k-1)\times
(k-1)$ matrix, which is the same as in \citet{Bourgain}.
Here, let $\m^*=(\mu_1,\ldots,\mu_{k-1})$ be the $(k-1)$-vector such
that $\m=\m^*\otimes\1$ under the null
hypothesis. We show that, under the null hypothesis, the
quasi-likelihood estimate of $\m^*$ is given by
%
%e2.10 ###
\begin{equation}
\hat{\m}^*=(\hat{\mu}_1,\ldots,\hat{\mu}_{k-1})'=(\1'\R^{-1}\1
)^{-1}\bigl(\mathbf{I}_{k-1}\otimes(\1'\R^{-1})\bigr)\Y,
\end{equation}
where $\mathbf{I}_{k-1}$ is a $(k-1)\times(k-1)$ identity matrix.
Thus, $\hat{\m}=\hat{\m}^*\otimes\1$. We obtain an explicit form of
the generalized quasi-likelihood scoring statistic~as
%
%e2.11 ###
\begin{equation}
W_G=C\cdot(\Y-\hat{\m})'\bigl(\hat{\F}^{-1}\otimes(\R^{-1}\X\X'\R
^{-1})\bigr)(\Y-\hat{\m}),
\end{equation}
where $C=[\X'\R^{-1}\X-\X'\R^{-1}\1(\1'\R^{-1}\1)^{-1}(\1'\R
^{-1}\X)]^{-1}$ is a constant depending on the trait
vector $\X$ and the correlation matrix $\R$, and $\hat{\F}$ is
computed by using the $\hat{\m}^*$. See
Appendix~C in the supplementary material for derivations of $\hat{\m}
^*$ and $W_G$ in the multiallelic case [\citet{Feng}]. Under the null
hypothesis, $W_G$
follows an $\chi^2$ distribution with $k-1$ degrees of freedom
asymptotically. Alternatively, we can express the
statistic in the form
%
%e2.12 ###
\begin{equation}
W_G=C\sum_{j=1}^{k-1}\sum_{l=1}^{k-1}(\hat{\F}^{-1})_{jl}(\Y
_j-\hat{\mu}_j\1)'\R^{-1}\X\X'\R^{-1}(\Y_l-\hat{\mu}_l\1).
\end{equation}
In the biallelic case that $k=2$, we have
$\F=\frac{1}{2}\mu(1-\mu)$ and $\Sig=\frac{1}{2}\mu(1-\mu)\R$,
$\hat{\m}^*$ and $W_G$ reduce to those that
are derived under the biallelic case. When the $n$ individuals in the
sample comprise subjects that are from $F$
independent families, we retain the notation of $\X_f, \1_f$ and $\R
_f$ as in the biallelic case. Let
$\Y_f=(\Y_{1f}',\ldots,\Y_{k-1,f}')'$ and $\Y_{jf}=(Y_{j1},\ldots,
Y_{jn_f})'$. The statistic $W_G$ is given by
%
%e2.13 ###
\begin{eqnarray}
W_G&=&C\cdot
\sum_{j=1}^{k-1}\sum_{l=1}^{k-1}(\hat{\F}^{-1})_{jl} \nonumber
\\[-8pt]
\\[-8pt]
&&\hphantom{C\cdot
\sum_{j=1}^{k-1}\sum_{l=1}^{k-1}}{}\times\Biggl\{\sum
_{f=1}^F(\Y_{jf}-\hat{\mu}_j\1_f)'\R_f^{-1}\X_f \sum_{f=1}^F(\Y
_{lf}-\hat{\mu}_l\1_f)'\R_f^{-1}\X_f  \Biggr\},
\nonumber
\end{eqnarray}
where
$C= \{\sum_{f=1}^F\X_f'\R_f^{-1}\X_f-(\sum_{f=1}^F\X_f'\R
_f^{-1}\1_f)^2(\sum_{f=1}^F\1_f'\R_f^{-1}\1_f)^{-1} \}^{-1}$.
Under the null hypothesis, $W_G$ follows an $\chi^2_{k-1}$
distribution asymptotically.

%s2.3 ###
\subsection{Data collected from multiple subpopulations}\label{sec2.3}
In this paper we extend our GQLS method to a solution that overcomes
the problem of population stratification. Suppose a sample is collected
from $S$ different subpopulations, denoted by $\mathit{pop}_1,\ldots, \mathit{pop}_S$. For
illustration, let the marker of interest be bi-allelic (e.g., an SNP).
For each subpopulation, $\mathit{pop}_s$, we compute a~GQLS test statistic,
$W_G^{(s)}$.\vspace*{-2pt} We know that the $W_G^{(s)}$ follows $\chi_1^2$
distribution asymptotically. In statistical theory, the sum of $S$
independent $\chi^2$ random variables follows an $\chi^2$
distribution with the degrees of freedom being the sum of the~$S$
degrees of freedom. Thus, a new overall statistic, which is the sum
over all subpopulations, having the form as
\[
W_{\mathit{all}}=W_G^{(1)} +W_G^{(2)} +\cdots + W_G^{(S)}
\]
follows an $\chi_S^2$ distribution asymptotically under the null hypothesis.

It is well known that FBAT is robust to the analysis of family data
collected from different populations. We will compare the performance
of our overall test method with FBAT in the population stratification
problem via simulation studies. We will also apply this overall test
method to the COGA data set. See Sections~\ref{sec3.3} and~\ref{sec4.2} for details.

%s3 ###
\section{Simulation study}\label{sec3}
We conduct simulation studies to validate the $\chi^2$ distribution
approximation to the distribution of the
$W_G$ statistic and to compare the power achieved by our approach with
the power achieved by the FBAT. We
consider three different study designs. First, we simulate single large
complex pedigrees. Second, we simulate multiple small families. Third,
for each study design, we combine samples simulated under settings to
mimic a sample collected from different subpopulations to investigate
the robustness of our extended method using the $W_{\mathit{all}}$ statistic.
Since SNPs are
popular for genetic association studies and SNPs are typically
biallelic, we simulate biallelic markers for demonstration. We use the
software KinInbcoef [\citet{KinInbcoef}] to compute the
kinship-inbreeding coefficient correlation matrix $\R$. We will
describe the simulation procedures and summarize the results for each
design in the following three subsections.

%s3.1 ###
\subsection{Single large pedigree study design}\label{sec3.1}
In this study design a family is grown starting from a single
individual. Each single individual is assigned
a~spouse with probability 0.8 or remains single with probability 0.2.
For each couple, we generate the number of offspring according to a
Poisson distribution with mean 3. Any pedigree that stops growing
before the completion of six generations by natural degeneration, or
stops before reaching to a desired family size, is disregarded. A new
pedigree is grown until we obtain one single pedigree that consists of
six generations and has
a desirable number of family members in the last three generations. In
our simulation study, we generate three large single outbred pedigrees
that have sizes of 136, 273, and 557, respectively. Family members of
the top three generations are removed to mimic the practical situations
(especially in human data) in which clinical information and DNA
samples are most likely not available for more than three generations
back. The genealogy of the entire pedigree remains for calculating the
correlation matrix $\R$. Removing the family members from the top
three generations, the pedigree sizes reduce to 124, 251, and 526,
respectively. For each founder (an individual with parents' genetic
information unknown), the marker genotype is simulated by random
mating. The genotypes of descents are generated according to the
Mendelian law of segregation.

To assess the type I error rate, for each individual, traits are
generated genetically according to an SNP with the minor allele
frequency (MAF) of the SNP being set to 0.3.
Denote the genotype of the SNP by $G$ that $G =0, 1,$ or 2 for having
0, 1, or 2 allele 1 in the genotype.
We simulated the quantitative trait, $X$, from $N(-1+G, \sigma^2)$
with $\sigma=1.2$. The binary trait was simulated from
Bernoulli($p_G$) with $p_0=0.1, p_1=0.3, p_2=0.4$.
Then, an SNP that is unlinked to the causal SNP is generated. The minor
allele frequency of the SNP is set to 0.3 and 0.1. For each combination
of settings, we generate $1\mbox{,}000$ replicates. For each simulated
data set, we compute the~$W_G$ statistic for the unlinked SNP, and take
the rejection threshold to be the $(1-\alpha)$th quantile of
the $\chi_1^2$ distribution. We run FBAT on each simulated data set.
In FBAT, default options are chosen
in most of the cases except that the ``minsize'' (the minimum number of
informative families) is set to 4. To illustrate the preservation of
the type I error by considering the correlation among related
subjects, we perform the standard Armitage trend test [\citet{Armitage}] that assumes independent subjects in the sample. The
Armitage trend test was implemented using the ``independence\_test''
function in the R package ``coin'' [\citet{R}]. This function also
allows testing on the quantitative trait.
We consider $\alpha=0.05$ and 0.01. In
Table~\ref{tab1} we summarize the empirical rejection rates at each significance
level for each combination of settings. The simulation results
indicate that the $\chi_1^2$ distribution approximates the
distribution of the $W_G$ statistic well. The inflation of the null
empirical rejection rate using the trend test is obvious (indicated in
bolded numbers) in the single large pedigree study.

To compare the power with the FBAT method, we simulate the quantitative
trait and the binary trait conditioning
on the genotype of each individual. The minor allele frequency of the
association marker is set to 0.3 and 0.1. Three
different genetic models are considered for both the quantitative and
binary trait. The
quantitative trait $X$ is generated according to an additive model:
$X_i=a+bG_i+\varepsilon_i$, where
\[
G_i =
\cases{\displaystyle
-1 ,&\quad for the homozygous genotype that  $Y_i=0$,\cr\displaystyle
0 ,&\quad for the heterozygous genotype that  $Y_i=1/2$,\cr\displaystyle
1 ,&\quad for the homozygous genotype that  $Y_i=1$.
}
\]
The random environmental errors $\varepsilon_i$, are generated from
$N(0, \sigma^2)$.
Without loss of generality, we set the intercept
$a=0$. We specify three different association models: (1) $b=0.5,
\sigma=1.2$; (2) $b=1, \sigma=1.5$; and (3)
$b=1, \sigma=1.2$. The coefficient $b$ quantifies the effect of the
marker. The different values of~$\sigma^2$
pose different levels of difficulty for the detection of genetic
association. These three models are denoted by qt1, qt2, and qt3,
respectively, in the tables that summarize the results of power assessments.

%
%t1 ###
\begin{table}
\tabcolsep=0pt
\caption{Type I error assessment---single large pedigree study design (6
generations)}
\label{tab1}\begin{tabular*}{\textwidth}{@{\extracolsep{\fill}}lccccccccccc@{}}
\hline
& & &\multicolumn{9}{c@{}}{\textbf{Sample size}} \\[-5pt]
& & &\multicolumn{9}{c@{}}{\hrulefill}\\
& & &\multicolumn{3}{c}{\textbf{124}} & \multicolumn{3}{c}{\textbf{251}} &\multicolumn{3}{c@{}}{\textbf{526}}\\[-5pt]
& & &\multicolumn{3}{c}{\hrulefill} & \multicolumn{3}{c}{\hrulefill}
&\multicolumn{3}{c@{}}{\hrulefill}\\
\textbf{MAF}& $\bolds\alpha$\tsup{1} & \textbf{Trait} & \textbf{GQLS} &\textbf{FBAT} &\textbf{Trend}& \textbf{GQLS}
&\textbf{FBAT}&\textbf{Trend} & \textbf{GQLS} &\textbf{FBAT} &\textbf{Trend} \\
\hline
0.3 & 0.05 & bt\tsup{2}& 0.049 & 0.051 &\textbf{0.086}&
0.045& 0.049 &\textbf{0.103}& 0.049 & 0.051&0.069 \\
& & qt\tsup{3}& 0.048& 0.046&\textbf{0.152}& 0.057 & 0.048
&\textbf{0.11}\hphantom{0} & 0.053 & 0.051 &\textbf{0.107}\\[3pt]
& 0.01 & bt & 0.006 & 0.007 &\textbf{0.019}& 0.009 & 0.009&\textbf
{0.033} & 0.011 & 0.011 &0.015\\
& & qt & 0.015 & 0.011 &\textbf{0.069}& 0.009 & 0.008&\textbf{0.03}\hphantom{0}&
0.013 & 0.01\hphantom{0} &\textbf{0.034}\\[6pt]
0.1 & 0.05 & bt & 0.052 & 0.076 &0.048 & 0.047 & 0.038 &0.053 & 0.048 &
0.048 &0.065\\
& & qt & 0.045 & 0.057 &\textbf{0.086}& 0.054 & 0.051 &0.059 & 0.054 &
0.044 &0.062\\[3pt]
& 0.01 & bt & 0.013 & 0.004&0.009& 0.009 & 0.006 &0.007 & 0.013 & 0.012
&\textbf{0.018}\\
& & qt & 0.014 & 0.008 &0.016& 0.012 & 0.012 &0.017& 0.011 & 0.009
&0.008\\
\hline
\end{tabular*}
\legend{{\tsup{1}Monte
Carlo standard deviation${}={}$0.0069 or 0.0031 for $\alpha= 0.05$ or
0.01, respectively.}
{\tsup{2}bt: binary
trait.}
{\tsup{3}qt: quantitative trait.}}
\end{table}
%

%
%t2 ###
\begin{table}
\tabcolsep=0pt
\caption{Power comparison---single large family study design (6 generations)}
\label{tab2}\begin{tabular*}{\textwidth}{@{\extracolsep{\fill}}lccd{1.3}d{1.3}d{1.3}d{1.3}d{1.3}d{1.3}@{}}
\hline
& & &\multicolumn{6}{c@{}}{\textbf{Sample size}}\\[-5pt]
& & &\multicolumn{6}{c@{}}{\hrulefill}\\
& & & \multicolumn{2}{c}{\textbf{124}} & \multicolumn{2}{c}{\textbf{251}}&\multicolumn{2}{c@{}}{\textbf{526}}\\[-5pt]
& & & \multicolumn{2}{c}{\hrulefill} &
\multicolumn{2}{c}{\hrulefill}&\multicolumn{2}{c@{}}{\hrulefill}\\
\textbf{MAF}& \textbf{Trait}& $\bolds\alpha$ & \multicolumn{1}{c}{\textbf{GQLS}} & \multicolumn{1}{c}{\textbf{FBAT}}
& \multicolumn{1}{c}{\textbf{GQLS}} & \multicolumn{1}{c}{\textbf{FBAT}} & \multicolumn{1}{c}{\textbf{GQLS}} &\multicolumn{1}{c@{}}{\textbf{FBAT}} \\
\hline
0.3&bt1 & 0.05&0.228 &0.186&0.368 &0.277 &0.603 &0.466\\
&& 0.01 &0.106 &0.06&0.186 &0.106 &0.386 &0.257\\[3pt]
&bt2 &0.05 &0.421 &0.264&0.663 &0.451 &0.928 &0.764 \\
& & 0.01 &0.218 &0.089&0.444 &0.237 &0.791 &0.548\\
[3pt]
&bt3 & 0.05 &0.610 &0.421& 0.91 &0.705 &0.997 &0.96 \\
& & 0.01 &0.385 &0.181& 0.758 &0.439 &0.868 &0.851 \\
[6pt]
0.1&bt1 & 0.05& 0.098&0.063&0.103 &0.076 &0.135 &0.099\\
&& 0.01 &0.03 &0.01&0.03 &0.016 & 0.053 &0.027\\[3pt]
&bt2 &0.05 & 0.164&0.116 &0.211 &0.132 &0.294& 0.187\\
& & 0.01 & 0.063&0.029&0.088 &0.036 &0.145 &0.77\\[3pt]
&bt3 & 0.05 &0.452 &0.33& 0.624 &0.424 &0.911 &0.721 \\
& & 0.01 &0.263 &0.119& 0.442 & 0.167 &0.804 &0.463 \\
[6pt]
0.3& qt1 & 0.05& 0.468&0.416& 0.781 & 0.734 & 0.977 &0.96\\
&& 0.01 & 0.242& 0.19 & 0.572 & 0.491& 0.905& 0.869 \\[3pt]
&qt2 &0.05 & 0.831&0.791 & 0.991 &0.970 & 1 & 1 \\
& & 0.01 & 0.656 &0.558 & 0.946 & 0.897& 0.999 & 0.999\\[3pt]
&qt3 & 0.05 & 0.943 & 0.909 & 0.999 & 0.994 & 1 & 1 \\
& & 0.01 & 0.836&0.758 & 0.995&0.981 & 1 & 1 \\
[6pt]
0.1&qt1 & 0.05& 0.261&0.228&0.401 &0.364 &0.707 &0.650\\
&& 0.01 & 0.126&0.074&0.214 &0.173 &0.483 &0.403\\[3pt]
&qt2 &0.05 & 0.511&0.451&0.730 &0.674 &0.968 &0.944 \\
& & 0.01 &0.326 &0.23& 0.54 &0.428 &0.916 &0.842\\[3pt]
&qt3 & 0.05 & 0.658&0.602&0.868 & 0.82 &0.994 &0.979 \\
& & 0.01 & 0.469&0.349& 0.736& 0.642 & 0.978&0.922 \\
\hline
\end{tabular*}
\vspace*{-2pt}
\end{table}

For the binary trait, we generate the affection status of individuals
according to three disease
models. In model 1 we consider a recessive epistasis disease controlled
by two SNPs that are unlinked to each other. Individuals
having two copies of allele 1 at both SNPs have a penetrance [defined as
$f=\p(\mbox{affected}|\mbox{genotype})$] of $f_1=0.5$. Individuals
having two copies of allele 1 at one SNP but
not at the other SNP have a penetrance of $f_2=0.4$. Individuals with
fewer than two copies of allele 1 at both
SNPs have a penetrance of $f_3=0.1$. In model 2 we consider a dominant
epistasis disease controlled by two
SNPs that are unlinked\vadjust{\goodbreak} to each other. Individuals with at least one
copy of allele 1 at both SNPs have a penetrance
of $f_1=0.5$. All other individuals have a penetrance of $f_2=0.1$. In
model 3 we consider a single disease locus model with $f_1=0.5$ if an
individual has two allele 1's at the SNP,
$f_2=0.3$ if an individual has one allele 1 at the SNP, and $f_3=0.1$
otherwise. These three models are denoted by bt1, bt2, and bt3,
respectively, in the tables that summarize the results of power assessments.

For each combination of settings, we generate $1\mbox{,}000$ replicates. For each
simulated data set, we compute the $W_G$ and obtain the $p$-value by the
$\chi_1^2$ approximation. We run FBAT on each simulated data set. The
proportions of $p$-values $\leq\alpha$ are reported in Table~\ref{tab2}. Simulation
results show that our method outperforms the FBAT for a higher detection power
in all scenarios. Results are particularly striking for the binary
trait with small sample size.

We extend our simulation studies to a single pedigree that consists of
nine generations. Genotypes and clinical information of family members
in the top six generations are removed. The genealogy of the entire
pedigree remains for calculating the correlation matrix $\R$. We
generate two single large pedigrees having sizes of 704 and 875,
respectively. After removing the family members in the top six
generations, there are 615 and 795 individuals remaining. Similarly, we
set the MAF of 0.3 and 0.1. The results of type I error and power
assessments are summarized in Tables~1 and~2 in the supplementary
material [\citet{Feng}]. The simulation results are consistent to the
results of the studies with six generations. The empirical type~I error
rates obtained by our method and the FBAT are close to each of the
nominal significance levels. The trend test generally inflates the
empirical rejection rate under the null hypothesis (indicated in bolded
numbers). Our method is generally more powerful than the FBAT.

%
%t3 ###
\begin{table}
\tabcolsep=0pt
\caption{Type I error assessment---multiple families study design}
\label{tab3}\begin{tabular*}{\textwidth}{@{\extracolsep{\fill}}lccccccccccc@{}}\hline
& & &\multicolumn{9}{c@{}}{\textbf{Sample size}} \\[-5pt]
& & &\multicolumn{9}{c@{}}{\hrulefill}\\
& & &\multicolumn{3}{c}{\textbf{100}} & \multicolumn{3}{c}{\textbf{200}} &\multicolumn{3}{c@{}}{\textbf{500}}\\[-5pt]
& & &\multicolumn{3}{c}{\hrulefill} & \multicolumn{3}{c}{\hrulefill}
&\multicolumn{3}{c@{}}{\hrulefill}\\
\textbf{MAF}& $\bolds\alpha$& \textbf{Trait} &\textbf{GQLS} &\textbf{FBAT}&\textbf{Trend}& \textbf{GQLS}&\textbf{FBAT}
&\textbf{Trend} &\textbf{GQLS}
&\textbf{FBAT}&\textbf{Trend} \\
\hline
0.3 & 0.05 & bt & 0.055 & 0.037&\textbf{0.1}\hphantom{000}& 0.048& 0.053 &\textbf
{0.09}\hphantom{00}& 0.056 & 0.051&0.057\\
& & qt & 0.056 & 0.058&0.0654 & 0.052 & 0.049 &0.07\hphantom{00}& 0.055 &
0.054&0.064\\[3pt]
& 0.01 & bt & 0.012 & 0.005 &\textbf{0.025}\hphantom{0}& 0.012 & 0.010 &0.018\hphantom{0}&
0.012 & 0.013&0.012 \\
& & qt & 0.013& 0.010 &0.009\hphantom{0}& 0.013 & 0.008 &\textbf{0.020}\hphantom{0}& 0.011 &
0.011&0.015 \\[6pt]
0.1 & 0.05 & bt & 0.054 & 0.037 &0.059\hphantom{0}& 0.050 & 0.045 &0.0068& 0.05\hphantom{0} &
0.05\hphantom{0} &0.065\\
& & qt & 0.048 & 0.043 &\textbf{0.082}\hphantom{0}& 0.047& 0.048 &\textbf{0.088}\hphantom{0}&
0.043 & 0.055&0.070\\[3pt]
& 0.01 & bt & 0.015 & 0.006 &0.011\hphantom{0}& 0.01\hphantom{0} & 0.009&0.013\hphantom{0} & 0.007 &
0.006&0.013 \\
& & qt & 0.013 & 0.007 &\textbf{0.022}\hphantom{0}& 0.007 & 0.006&\textbf{0.031}\hphantom{0}
& 0.007 & 0.006 &0.013\\ \hline
\end{tabular*}
\end{table}
%

%s3.2 ###
\subsection{Multiple families study design}\label{sec3.2}
In this study families are grown following the similar procedure as for
the single large family
study design except that families will grow for a maximum of three
generations. The simulated sample comprises families and independent
individuals. Family sizes range from 1 to 23 with an average size of
6.3. As in the single large pedigree study design, the genotype of
founders is generated by random mating and the genotype of nonfounders
is generated according to the Mendelian law of segregation. We let the
sample size (number of subjects) be 100, 200, and 500, respectively.
To assess the type I error rate, we
generate a quantitative trait and a binary trait for each individual as
described in the single large family study design. Then, an SNP that is
unlinked to the causal SNP is generated. The minor allele frequency of
the SNPs is set to 0.3 and 0.1. For each combination of settings, we
generate $1\mbox{,}000$ replicates. In
Table~\ref{tab3} we summarize the null empirical rejection rates. The results
indicate that the~$\chi_1^2$ distribution approximates the
distribution of the $W_G$ statistic well. The inflation of the null
empirical rejection rate using the trend test is observed.
For power comparisons, we simulate the quantitative traits and binary
traits according to the six models that
have been described in the previous section. The MAF of the association marker
is also set to 0.3 and 0.1. The powers achieved by our method and
the\vadjust{\eject}
FBAT under each combination of settings are
summarized in Table~\ref{tab4}. Simulation results show that our method
consistently outperforms FBAT for all scenarios.

%
%t4 ###
\begin{table}
\tabcolsep=0pt
\caption{Power comparison---multiple small pedigree study design}
\label{tab4}\begin{tabular*}{\textwidth}{@{\extracolsep{\fill}}lccd{1.3}d{1.3}d{1.3}d{1.3}d{1.3}d{1.3}@{}}
\hline
& & &\multicolumn{6}{c@{}}{\textbf{Sample size}}\\[-5pt]
& & &\multicolumn{6}{c@{}}{\hrulefill}\\
& & & \multicolumn{2}{c}{\textbf{100}} & \multicolumn{2}{c}{\textbf{200}} &\multicolumn{2}{c@{}}{\textbf{500}}\\[-5pt]
& & & \multicolumn{2}{c}{\hrulefill} & \multicolumn{2}{c}{\hrulefill}
&\multicolumn{2}{c@{}}{\hrulefill}\\
\textbf{MAF}& \textbf{Trait}& $\bolds\alpha$ & \multicolumn{1}{c}{\textbf{GQLS}} & \multicolumn{1}{c}{\textbf{FBAT}} & \multicolumn{1}{c}{\textbf{GQLS}} & \multicolumn{1}{c}{\textbf{FBAT}} & \multicolumn{1}{c}{\textbf{GQLS}} &\multicolumn{1}{c@{}}{\textbf{FBAT}} \\
\hline
0.3&bt1 & 0.05&0.22 &0.17&0.302&0.171&0.610&0.388\\
&& 0.01 &0.093&0.04&0.135&0.07&0.402&0.215\\[3pt]
&bt2 &0.05 &0.354 &0.178&0.561&0.317&0.93&0.675\\
& & 0.01 &0.170 &0.044&0.339&0.11&0.818&0.42\\
[3pt]
&bt3 & 0.05 &0.639&0.328&0.829&0.514&0.996&0.917 \\
& & 0.01 &0.38&0.13&0.643&0.255&0.982&0.78 \\
[6pt]
0.1&bt1 & 0.05&0.095&0.063&0.095&0.082&0.118&0.073\\
&& 0.01 &0.029&0.008&0.041&0.01&0.032&0.011\\[3pt]
&bt2 &0.05 &0.132&0.078&0.183&0.107&0.322&0.16 \\
& & 0.01 &0.046&0.011&0.078&0.024&0.148&0.06\\
[3pt]
&bt3 & 0.05 &0.118&0.073&0.322&0.16&0.942&0.634 \\
& & 0.01 &0.032&0.011&0.148&0.06&0.843&0.361\\
[6pt]
0.3& qt1 & 0.05&0.433&0.309&0.709&0.546&0.98&0.928\\
&& 0.01 &0.217 &0.114&0.478&0.302&0.921&0.777 \\[3pt]
&qt2 &0.05 &0.795 &0.624&0.969&0.849&1&1\\
& & 0.01 &0.58 &0.369&0.913&0.704&1&1\\
[3pt]
&qt3 & 0.05 &0.934&0.792&0.999&0.976&1&1\\
& & 0.01 &0.817 &0.552&0.99&0.691&1&1\\
[6pt]
0.1&qt1 & 0.05&0.215 &0.153&0.351&0.264&0.746&0.597\\
&& 0.01 &0.079&0.046&0.157&0.091&0.520&0.331\\[3pt]
&qt2 &0.05 &0.456 &0.279&0.734&0.502&0.986&0.726 \\
& & 0.01 &0.228&0.09&0.515&0.251&0.943&0.796\\
[3pt]
&qt3 & 0.05 &0.647 &0.391&0.895&0.625&0.99&0.977 \\
& & 0.01 &0.419 &0.149&0.745&0.4&0.992&0.916\\
\hline
\end{tabular*}
\vspace*{-2pt}
\end{table}
%

%s3.3 ###
\subsection{Data with subpopulations}\label{sec3.3}
In this section we consider the situation that a sample contains
individuals from different populations. Similarly to the previous
section, we consider biallelic markers. For illustration, we consider a
sample collected from two subpopulations only. In fact, for each of the
previous study designs, the single large pedigree and the multiple
small pedigrees, we combine two simulated data sets with different MAF
to make up a sample that consists of individuals from two different
populations. For example, in the single large pedigree study design, we
combined the two simulated samples from two subpopulations with MAF
being set to 0.1 and 0.3, and with different combinations of sample
sizes for each subpopulation. For each combined sample, the $W_{\mathit{all}}$
is the sum of the two $W_G$ statistics from two subsamples. The
$p$-values are obtained by the $\chi_2^2$ distribution. The type~I error
rate and the power are compared between our method and FBAT.

In the supplementary material, Table~3, we summarize the results of
type~I error rates assessment by combining two single large pedigrees
[\citet{Feng}]. Similarly, in the supplementary material, Table~4, we
summarize the results of type~I error assessment by combining the two
simulated samples of multiple small pedigrees [\citet{Feng}].
Overall, the empirical type~I error rates obtained by our method using
the $W_{\mathit{all}}$ test statistics and the empirical type~I error rates
obtained by FBAT are close to each of the nominal significance levels.
However, FBAT is slightly less stable. For example, in Table~\ref{tab3}, the
empirical type~I error rate is 0.005 at 0.01 significance level for
a~quantitative trait when combining the sample size of 124 from
population~1 and sample size of 526 from the population 2. In Table~\ref{tab4},
the empirical error rate is 0.033 at 0.05 significance level for a
binary trait when combining the sample sizes of 100 from both
population~1 and population 2. Both of the 95\% confidence intervals
constructed based on these two empirical type~I error rates do not
cover the true values of $\alpha= 0.01$ and 0.05.

In the supplementary material, Tables~5 and~6, we summarize the results
of power assessment [\citet{Feng}]. The simulation results indicated
that the performance of our method and FBAT are comparable that one
shows some advantages over the other under some suituations, and
vice versa.

%s4 ###
\vspace*{-2pt}
\section{Real data analysis}
\vspace*{-2pt}
\label{sec4}

%s4.1 ###
\subsection{Application to Holstein cattle data}\label{sec4.1}
The data set contains 821 proge\-ny-tested proven bulls born between 1965
and 2001. Each bull was genotyped using
the Affymetrix MegAllele GeneChip Bovine mapping 10K SNP array [\citet{affy}]. Among 821 bulls, some bulls also
appear as the sires of other bulls. The relationships among bulls and
their sires and dams are complicated. All
of the 821 bulls sampled have genetically contributed to the current
Canadian cow population. Most of the animals
in the population have a nonzero inbreeding coefficient. A genealogy of
the population tracing back 25
generations, with the oldest animal born in 1909, was used to compute
the kinship-inbreeding coefficient
with the software CFC [\citet{CFC}]. Out
of 9,919 genotyped SNPs, only 8,624 SNPs
have known location on the 29 Bos Taurus autosome chromosomes (BTA).
SNPs with more than 20\% of missing values
or MAF of less than 5\% were excluded from the study. A total of $7\mbox{,}103$
SNPs were analyzed. The experimental
design is mainly a granddaughter design that the milk productivities of
daughters and granddaughters of a bull\vadjust{\eject}
are used to estimate the breeding value of the bull. The phenotypes
used in the analysis were trait EBVs
released in November 2008 and provided by the Canadian Dairy Network
(CDN, Guelph, Canada). For illustration, we
only present results of the association tests with milk yield EBV.

In Table~\ref{tab5} we report the top 81 most significant SNPs that have
$p$-va\-lue${}\leq{}$0.001 that can be grouped into 36
regions (SNPs at a close inter-distance, less than 1cM, define a
region) on 16 BTAs.
Out of 36 significant SNPs
or regions, 16 significant SNPs or regions have been found in agreement
with the quantitative traits loci or
associated SNPs reported in the literature. In BTA14, 22 SNPs
concentrated in 0--27cM have strong association
with milk yield and their $p$-values range from $6.45\times10^{-10}$
to 0.001. At the telomere of BTA14,
Daicylglycerol acyl transferase 1 (\textit{DGAT1}) at 0cM has been
considered to be a quantitative trait
nucleotide with a major effect on milk yield [\citet{Bennewitz}; \citet{Boichard};
\citet{Grisart}]. An SNP at 0.27cM has a strong
association signal. Twelve SNPs in the region of 3.38--8.47cM are
consistent with 3
SNPs at 4cM, 5cM, and 6cM that have been reported significantly
associated with milk yield by \citet{Daetwyler}
and \citet{Bennewitz}. An SNP at 11.2cM also confirms the association
with milk yield reported by
\citet{Daetwyler}. The most significant SNP is found at 94cM on BTA5 and
confirms a QTL at the same location reported by \citet{Viitala03}. A
significant SNP at 98cM also confirms a QTL at the same location
reported by \citet{Viitala03}.
Note that, after adjusting for Bonferroni's correction at 5\%
significance level (or at 7.13$\times10^{-6}$ individual significance level),
11 regions remain significant. However, for many complex traits that
are controlled by several genes, each individual gene may only have a
small effect. When thousands of SNPs are tested, using the Bonferroni's
correction may result in low power of the study.
Therefore, when we interpret the Bonferroni result, we need to be
careful that some signals disappearing after the adjustment may be due
to the conservativeness of Bonferroni's correction.

%
%t5 ###
\begin{table}[t!]
\tabcolsep=0pt
\caption{Most significant loci ($p$-value${}\leq{}$0.001) found for milk yield
trait}
\label{tab5}\begin{tabular*}{\textwidth}{@{\extracolsep{\fill}}ld{2.0}cc@{}}
\hline
\textbf{BTA} & \multicolumn{1}{c}{\textbf{No. of SNPs}} & \textbf{Location (cM)}\tsup{1}
&$\bolds p$\textbf{-value}\tsup{2}\\
\hline
\hphantom{2}1& 1 & 47.90\tsup{5} & 2.18$\times10^{-5}$\\
\hphantom{2}4 & 1 & 20.05\tsup{5} & 0.000105\\
& 2 &56.65, 59.81\tsup{5}&0.000664\\
&1 & 101.74 & 0.000126\\
\hphantom{2}5 &1&1.03\tsup{5}&0.00086\\
& 1 &8.32 & $2.91\times10^{-5}$\\
&3 &29.59--34.46 & $7.5\times10^{-5} $ \\
& 6 & 45.51--50.53 & $4.65\times10^{-6} $\tsup{*}\\
&1 & 69.89 & $8.8\times10^{-6}$\\
& 8 & 73.49--77.77 &$8.44\times10^{-7} $\tsup{*}\\
& 12 &90.76--101.06\tsup{6,8,9}& $3.14\times10^{-11} $\tsup{*}\\
& 1 & 114.90\tsup{3} & 0.000125\\
\hphantom{2}6 & 1 &47.66\tsup{7}& 0.000355\\
\hphantom{2}7 & 1 & 75.07\tsup{5} & 0.001\\
\hphantom{2}8 &1 & 41.75 & 0.000215\\
&1 &55 & 0.000126\\
11 & 1&113.46 & 0.000853\\
12 &1 & 61.77\tsup{5} & $5.81\times10^{-7} $\tsup{*}\\
14 & 1 & 0.27\tsup{3,4,5,6}& $2.06\times10^{-6} $\tsup{*}\\
& 12 &3.38--8.47\tsup{3,5} & $3.86\times
10^{-8} $\tsup{*}\\
&3 & 11.2\tsup{4,5}&$4.94\times10^{-6}
$\tsup{*}\\
&4 & 21.50\tsup{4}& $6.45\times10^{-10} $\tsup{*}\\
&2 &26.69 & 0.000691\\
15 & 1 & 21 & 0.000291\\
16 & 1 & 31.66 & $3.39\times10^{-8}$\\
& 1 &54 & 0.000364\\
&1 &62 &0.000946\\
&1 &90.54\tsup{4} & $3.85\times10^{-5}$\\
17 & 1 &16 & 0.000595\\
&1 & 72& 0.00034\\
& 1 &78.58& 0.000868\\
18 & 1 &15.78\tsup{4} & $1.75\times10^{-6} $\tsup{*}\\
23 & 2 & 9.36 & 4.44$\times10^{-6} $\tsup{*}\\
26 & 2 &44, 45 & 0.000341\\
& 1& 53\tsup{3,4} & 0.00061\\
27 & 1 &57 & 0.000962\\
\hline
\end{tabular*}
\legend{{\tsup{1}Chromosomal region that the SNPs span on.}
{\tsup{2}Minimum
$p$-value if there is more that one SNP in the region.}
{\tsup{3}In
agreement with \citet{Bennewitz}.}
{\tsup{4}In
agreement with \citet{Boichard}.}
{\tsup{5}In
agreement with \citet{Daetwyler}.}
{\tsup{6}In
agreement with \citet{Grisart}.}
{\tsup{7}In
agreement with \citet{Heyen}.}
{\tsup{8}In
agreement with \citet{Viitala03}.}
{\tsup{9}In
agreement with \citet{Viitala08}.}
{\tsup{*}Significant at 5\% Bonferroni's
correction (at 7.13$\times10^{-6}$ individual significance level).}}
\end{table}
%

%s4.2 ###
\subsection{Application to COGA data}\label{sec4.2}
The Collaborative Study on the Genetics of Alcoholism (COGA) data set
was provided by the Genetic Analysis
Workshop 14 (GAW14). The data set included $1\mbox{,}614$ individuals from 143
families. Among $1\mbox{,}614$ individuals, 1,351
individuals were genotyped for a~panel of 11,555 SNPs from Affymetrix.
A set of alcoholism phenotypes and
covariates were provided. We use the ALDX1 as the phenotype.
Individuals who are coded as ``affected'' in the ALDX1 variable are
considered as affected individuals. Unaffected individuals are those
coded as ``pure'' unaffected in the ALDX1. Individuals with other
codings are considered to have unknown phenotypes. In this study, we
compare our method with FBAT under three scenarios. In scenario 1 we
consider a large sample from a single population. We only include
individuals who are coded as ``white, non-Hispanic.'' There are 119 such
families consisting of $1\mbox{,}074$ individuals.\vadjust{\eject} In scenario 2 we consider a
small sample from a single population. We only include individuals who
are coded as ``white, Hispanic.'' There are 11 such families consisting
of 78 individuals. In scenario 3 we combine the two samples from the
the two populations of ``white, Hispanic'' and ``white, non-Hispanic.''
In our studies, we use the software KinInbcoef to compute the kinship
coefficient for correlation matrix $\R$. We only analyze SNPs that are
on autosomes. In total, there are $10\mbox{,}532$ SNPs on autosomes.

The results based on our method are summarized in the supplementary
material, Table~7 [\citet{Feng}]. In total, there are 22 SNPs found to
be significant ($p$-values${}<{}$0.001) in the ``white, Hispanic'' sample, 19
SNPs are found to be significant based on the ``white, non-Hispanic''
sample, and 24 SNPs are found to be significant based on the pooled
samples of ``white.'' There are 19 SNPs that are significant in both the
pooled sample and the ``white, Hispanic'' or in both the pooled sample
and the ``white, non-Hispanic'' sample. On chromosome 2, SNP tsc0052826
is significant in both the ``white Hispanic'' sample and the pooled
sample; it is 0.344cM from a marker that had been reported for a
significant linkage with alcohol dependence [\citet{Hill}; \citet{Valdes}]. On
chromosome 6, SNP tsc1395926 is significant in both the ``white
Hispanic'' sample and the pooled sample. It is very close to two loci
(less than 1Mb) that had been found to link to the alcoholism [\citet{Hill}; \citet{Ma}]. On Chromosome 7, SNPs tsc0333356 is significant in both
the ``white Hispanic'' sample and the pooled sample; it is 1.47cM away
from a~marker that had been reported to significantly link to ALDX1 by
\citet{Zhu} and is 0.811cM from a marker that has shown significant
linkage to alcohol dependence by \citet{Hill}. The most significant
SNP is SNP tsc0059716 on chromosome 13 ($p$-value${}={}$4$\times10^{-6}$),
which is about 2.4cM away from an SNP that had been reported to
significantly associate with ALDX1 [\citet{Zhu}].
In total, there are 12 SNPs found to be very close to regions or SNPs
that had been reported to link or associate with alcohol dependence or
alcoholism related traits in the literature. After adjusting for
Bonferroni's correction at 5\% significance level (or at 4.75$\times
10^{-6}$ individual significance level), four SNPs (tsc0587314 on
chromosome~3, tsc0506913 on chromosome 5, tsc0630829 on chromosome 7,
and tsc0059716 on chromosome 13) remain significant.

The results based on FBAT are summarized in Table~8 in the
supplementary material [\citet{Feng}]. In total, there are 43 SNPs
found to be significant ($p$-value${}<{}$0.001) in the pooled sample, 29
SNPs are significant in the ``white, non-Hispanic'' sample, and only one
SNP is significant in the ``white, Hispanic'' sample.
Among these significant SNPs, SNP tsc0056748 on chromosome 13 is
significant in more that one sample (the pooled sample and the ``white,
non-Hispanic'' sample). There are 17 significant SNPs in the pooled
sample that had been reported significantly associated with the ALDX1
by \citet{Zhu}. Note that the\vadjust{\eject} results in \citet{Zhu} are based on the
same pooled sample of ``white, Hispanic'' and ``white, non-Hispanic''
same definition of ``affected'' individual, and are analyzed by the FBAT
as well. The only difference is the definition of ``unaffected''
individual, in that we only use ``pure-affected'' individuals while
\citet{Zhu} use ``pure-unaffected'' and ``never drank.'' Therefore,
there would be more significant SNPs confirmed by \citet{Zhu}. In
addition, SNP tsc0046578 on chromosome 1 is 1.37cM away from an SNP
that had been reported to significantly link to alcohol dependence by
\citet{Prescott}. SNP tsc0697701 on chromosome 8 is 0.7Mb away from an
SNP that significantly links to the alcoholism by \citet{Hill}. SNP
tsc0896393 on chromosome 12 is 1.5Mb away from an SNP that
significantly links to ALDX1 reported by \citet{Ma}. After adjusting
for the Bonferroni correction, three SNPs (tsc0515272 on chromosome 3,
tsc0029429 on chromosome 9, and tsc 1750530 on chromosome 16) remain
significant.

%s5 ###
\section{Discussion}\label{sec5}
In this article we adopt the framework of the generalized linear model
and assume that the expected marker
allelic frequency is connected to the linear predictor based on the
trait of interest through an arbitrary
specified link function. Although we focus on the logistic link, which
is the canonical link for
a binomial random variable, models utilizing other link functions can
be built with minor modifications of the
approach herein. The
population-based association study is still a popular study design for
common traits. To prevent spurious
association due to a confounding population structure, association
studies should be performed within a relative
homogeneous population. Such a population-based association study is
a~spe\-cial case of our method in which the
$\R$ matrix will be an identity matrix for independent subjects. For
the stratified
population, \citet{Lander06} suggested using ``internal controls'' to
balance the ethnicity between the cases
and controls in the sample in order to eliminate the confounding
effects. Our proposed generalized association
method uses all available family members to provide natural ``internal
controls.'' \citet{Conneally} pointed out that for any choice of study
design, whether based on families or
population-based, a large sample size is needed to detect an associated
gene with only a partial effect on the
trait. The quasi-likelihood scoring method fully utilizes the
correlation information among the sampled
individuals. It accommodates various
data types for genetic association studies including the conventional
population-based association studies, and
those using founder/isolated populations with documented genealogy, or
multiple complex pedigrees. Thus, this method
essentially increases the sample size and becomes more powerful.
On the other hand, when a data set contains samples from multiple
subpopulations, we propose a solution that combines the $W_G$
statistics from each subpopulation to construct a new test statistic~$W_{\mathit{all}}$.
The $W_{\mathit{all}}$ statistic is founded to follow an $\chi^2$
distribution asymptotically with the degrees of freedom depending on
the number of subpopulations and the number\vadjust{\eject} of alleles of the marker
being tested. Simulation results confirm that the $\chi^2$
distribution approximates the distribution of~$W_{\mathit{all}}$ well.
Simulation results also show that our method has comparable power to
the FBAT. However, our approach is limited to known subpopulations. If
unknown subpopulations exist, it is possible to extend our approach to
a mixture population with more population parameters to be estimated.

It is known that pedigree errors can easily arise in the study of large
pedigrees and even in the study of small pedigrees. Our GQLS method
cannot handle this error directly. However, many methods and software
are available to detect such errors under different study designs
[PREST by \citet{Sun}; RELATIVE by \citet{Goring}; RELPAIR by \citet{Epstein}]. When the pedigree errors are found, involved individuals
could be either removed from the study accordingly, or, the
relationship, that is, the kinship and inbreeding coefficients, among
involved individuals can be inferred through the genome scan (if genome
data are available) as a substitute in the $\R$ matrix. However, the
approximation of the $\chi^2$ distribution to the resulting $W_G$
statistic needs to be further investigated.

\section*{Acknowledgment}
The authors thank Professor Mary Thompson (Depart\-ment of Statistics and
Actuarial Science, University of
Waterloo) for a~criti\-cal reading of the original version of this paper.
The R code for computing the GQLS test statistic is available at
\href{http://www.uoguelph.ca/\textasciitilde zfeng/software/}{http://www.uoguelph.ca/\textasciitilde zfeng/}
\href{http://www.uoguelph.ca/\textasciitilde zfeng/software/}{software/}.

\begin{supplement}%[id=suppA]
\stitle{Mathematical justifications and additional results\\}
\slink[doi,text={10.1214/11-AOAS465SUPP}]{10.1214/11-AOAS465SUPP} %[doi,text={...}] - jei reikia
%suskaldyti doi
\slink[url]{http://lib.stat.cmu.edu/aoas/465/supplement.pdf}
\sdatatype{.pdf}
\sdescription{The supplementary materials of the paper are organized
as follows.
Appendix~A provides the theoretical justification of the
variance--covariance matrix $\Sig_0$.
Appendix~B derives the explicit form of the $W_G$ statistic for a
biallelic marker in a single
pedigree study design. Appendix~C derives the expression of the $W_G$
statistic for a~multi-allelic
marker in a single pedigree study design. In Appendix~D additional
results of simulation studies and
the results of COGA data analysis are summarized in tables.}
\end{supplement}

% imsref loaded by smiklovaite, 2011-04-27 15:13:05
% imsref loaded by smiklovaite, 2011-04-27 15:45:57
%

\printaddresses

\end{document}